\documentclass[12pt]{spieman}  % 12pt font required by SPIE;
\usepackage{amsmath,amsfonts,amssymb}
\usepackage{graphicx}
\usepackage{setspace}
\usepackage{tocloft}
\usepackage{color}
\title{Verification of the Timing System for the X-ray Imaging and Spectroscopy Mission in the GPS Unsynchronized Mode}

\author[a]{Megumi Shidatsu}
\author[b,c]{Yukikatsu Terada}
\author[d]{Takashi Kominato}
\author[b]{So Kato}
\author[b]{Ryohei Sato}
\author[b]{Minami Sakama}
\author[b]{Takumi Shioiri}
\author[b]{Yugo Motogami}
\author[a]{Yuuki Niida}
\author[a]{Chulsoo Kang}
\author[a]{Toshihiro Takagi}
\author[a]{Taichi Nakamoto}
\author[c]{Chikara Natsukari}
\author[b,c]{Makoto S. Tashiro}
\author[c]{Kenichi Toda}
\author[c]{Hironori Maejima}
\author[c]{Shin Watanabe}
\author[c]{Ryo Iizuka}
\author[c]{Rie Sato}
\author[e]{Chris Baluta}
\author[c]{Katsuhiro Hayashi}
\author[c]{Tessei Yoshida}
\author[c]{Shoji Ogawa}
\author[c]{Yoshiaki Kanemaru}
\author[c]{Kotaro Fukushima}
\author[c]{Akio Hoshino}
\author[f]{Hiromitsu Takahashi}
\author[g]{Masayoshi Nobukawa}
\author[f]{Tsunefumi Mizuno}
\author[h]{Kazuhiro Nakazawa}
\author[i]{Shin'ichiro Uno}
\author[c]{Ken Ebisawa}
\author[j]{Satoshi Eguchi}
\author[b]{Satoru Katsuda}
\author[k]{Aya Kubota}
\author[l]{Naomi Ota}
\author[m]{Atsushi Tanimoto}
\author[a]{Yuichi Terashima}
\author[n]{Yohko Tsuboi}
\author[o]{Yuusuke Uchida}
\author[p]{Hideki Uchiyama}
\author[l]{Shigeo Yamauchi}
\author[n]{Tomokage Yoneyama}
\author[q]{Satoshi Yamada}
\author[c]{Nagomi Uchida}
\author[e]{Matt Holland}
\author[r,e,s]{Michael Loewenstein}
\author[t,e,s]{Tahir Yaqoob}
\author[u]{Eric D. Miller}
\author[v]{Robert S. Hill}
\author[v]{Efrain C. Perez-Solis}
\author[v]{Morgan D. Waddy}
\author[v]{Mark Mekosh}
\author[v]{Joseph B. Fox}
\author[v]{Isabella S. Brewer}
\author[v]{Emily Aldoretta}
\author[e,s,t]{Koji Mukai}
\author[e,s,t]{Kenji Hamaguchi} 
\author[e,r]{Francois Mernier}
\author[e,r]{Anna Ogorzalek}
\author[e,s,t]{Katja Pottschmidt}
\author[e,w]{Mihoko Yukita}

\affil[a]{Ehime University, Graduate School of Science and Engineering, 2-5, Bunkyo-cho, Matsuyama-shi, Ehime, 790-8577, Japan}
\affil[b]{Saitama University, Graduate School of Science and Engineering, 255 Shimo-Ohkubo, Sakura-ku, Saitama-shi, Saitama, 338-8570, Japan}
\affil[c]{Japan Aerospace Exploration Agency (JAXA), Institute of Space and Astronautical Science, 3-1-1, Yoshinodai, Chuo-ku, Sagamihara-shi, Kanagawa, 252-5210, Japan}  
\affil[d]{NEC Corp., 1-10-2, Nishin-tyou, Fuchu-shi, Tokyo, 183-0036, Japan}
\affil[e]{National Aeronautics and Space Administration (NASA), Goddard Space Flight Center, 8800 Greenbelt Road, Greenbelt, Maryland, 20771, United States}
\affil[f]{Hiroshima University, School of Science, 1-3-2, Kagamiyama, Higashi-Hiroshima, Hiroshima, 739-0046, Japan}
\affil[g]{Nara University of Education, Department of Teacher Training and School Education, Takahata, Nara-shi, Nara, 630-8301, Japan}
\affil[h]{Nagoya University, Department of Physics, Furo-cho, Chikusa-ku, Nagoya-shi, Aichi, 464-8601, Japan}
\affil[i]{Nihon Fukushi University, Faculty of Health Sciences, 26-2, Higashi-Ikumi, Handa-shi, Aichi, 475-0012, Japan}
\affil[j]{Kumamoto Gakuen University, Faculty of Economics, 2-5-1, Oe, Chuo-ku, Kumamoto 862-8680, Japan}
\affil[k]{Shibaura Institute of Technology, Department of Electronic Information Systems, 307 Fukasaku, Minuma-ku, Saitama-shi, Saitama, 337-8570, Japan}
\affil[l]{Nara Women’s University, Department of Physics, Kitauoya-nishi, Nara-shi, Nara, 630-8506, Japan}
\affil[m]{Kagoshima University, Faculty of Science, 1-21-24, Kohrimoto, Kagoshima-shi, Kagoshima, 890-0065, Japan
}
\affil[n]{Chuo University, Faculty of Science and Engineering, Department of Physics, 1-13-27, Kasuga, Bunkyo-ku, Tokyo, 112-8551, Japan}
\affil[o]{Tokyo University of Science, 2641 Yamazaki, Noda-shi, Chiba, 278-8510, Japan}
\affil[p]{Shizuoka University, Faculty of Education, 836, Ohya, Suruga-ku,
Shizuoka-shi, Shizuoka, 422-8529, Japan}
\affil[q]{RIKEN, Nishina Center, 2-1 Hirosawa, Wako-shi, Saitama, 351-0198, Japan}
\affil[r]{University of Maryland, College Park, Maryland, 20742, United States}
\affil[s]{Center for Research and Exploration in Space Science and Technology (CRESST), NASA/GSFC, Greenbelt, Maryland, United States}
\affil[t]{University of Maryland, Baltimore County, Maryland, United States}
\affil[u]{Massachusetts Institute of Technology, Kavli Institute for Astrophysics and Space Research, 77 Massachusetts Avenue, Cambridge, Massachusetts, 02139, United States}
\affil[v]{ADNET Systems Inc., 6720B Rockledge Drive, Suite \# 504. Bethesda, Maryland, 20817, United States}
\affil[w]{Johns Hopkins University, 3400 N. Charles Street Baltimore, Maryland, 21218, United States}

\cftpagenumbersoff{figure}
\cftpagenumbersoff{table} 
\begin{document} 
\maketitle

\begin{abstract}
We report the results from the ground and on-orbit verifications of the XRISM timing system when the satellite clock is not synchronized to the GPS time. In this case, the time is determined by a free-run quartz oscillator of the clock, whose frequency changes depending on its temperature. In the thermal vacuum test performed in 2022, we obtained the GPS unsynchronized mode data and the temperature-versus-clock frequency trend. Comparing the time values calculated from the data and the true GPS times when the data were obtained, we confirmed that the requirement (within a 350 $\mu$s error in the absolute time, accounting for both the spacecraft bus system and the ground system) was satisfied in the temperature conditions of the thermal vacuum test. We also simulated the variation of the timing accuracy in the on-orbit temperature conditions using the Hitomi on-orbit temperature data and found that the error remained within the requirement over $\sim 3 \times 10^{5}$ s. The on-orbit tests were conducted in 2023 September and October as part of the bus system checkout. The temperature versus clock frequency trend remained unchanged from that obtained in the thermal vacuum test and the observed time drift was consistent with that expected from the trend. 
\end{abstract}

% Include a list of up to six keywords after the abstract

% Include email contact information for corresponding author
{\noindent \footnotesize\textbf{*}Megumi Shidatsu,  \linkable{shidatsu.megumi.wr@ehime-u.ac.jp} }

\begin{spacing}{1}   % use double spacing for rest of manuscript

\section{Introduction: XRISM Timing System and Requirement of Timing Accuracy}
\label{sect:intro}  % \label{} allows reference to this section

The X-ray Imaging and Spectroscopy Mission (XRISM)\cite{Tashiro2020,Tashiro2025}, which is an X-ray satellite developed by Japan Aerospace Exploration Agency (JAXA) and National Aeronautics and Space Administration (NASA) in collaboration with European Space Agency (ESA) and other international institutes, was successfully launched in 2023 September 7 JST 8:47. It carries the X-ray microcalorimeter Resolve\cite{Ishisaki2022}, which achieves an unprecedentedly high energy resolution of 7 eV (as designed) and the X-ray CCD Xtend\cite{Mori2022}, which provides wide-field ($38'.5 \times 38'.5$) X-ray imaging, installed at the focal planes of the X-ray Mirror Assemblies (XMAs). The scientific goals of the mission are to investigate the circulation of matter and transfer of energy in cosmic plasmas, and to explore the structure and evolution of the universe, through high-resolution spectroscopy combined with wide-field imaging capability. 
These goals include understanding the physics of high energy phenomena associated with compact objects, such as black holes and neutron stars. These objects often exhibit rapid variability on timescales as short as 1 ms, which carries crucial information about their nature and physical processes driving their activity. Resolve is designed to have timing capabilities sufficient to probe such rapid variability. To fully exploit this potential, it is essential to establish a sufficient timing accuracy across the entire satellite system.

As described in Terada et al. (2025)\cite{Terada2024JATIS}, the timing system of XRISM, consisting of the spacecraft itself and the ground system, 
adopts the same design as that of Hitomi\cite{Terada2018}. The source of the timing information is the quartz clock onboard the main computer of the spacecraft, named the Satellite Management Unit (SMU). 
It generates time digits named the Time Indicators (TIs) and provides them to the bus components and the scientific instruments: 
Resolve and Xtend, via the SpaceWire network. These instruments produce finer time information using their own clocks referring to the TIs from the SMU. The data taken on the satellite are delivered to the ground system, in which the TI values of the individual data are converted to the TIME values (the actual time in units of second from a reference time). 

XRISM carries the GPS receiver (GPSR), which synchronizes the TIs from the SMU clock to the International Atomic Time (TAI)\cite{tai1}\cite{tai2},
as explained in Terada et al. 2018\cite{Terada2018} and Terada et al. 2025\cite{Terada2024JATIS}. When the synchronization process works properly, it is straightforward to convert the TI values to the TIME values. 
The GPSR uses flight proven products and is expected to acquire the GPS signals continuously during the mission lifetime. Nevertheless, the SMU clock has a ``failure mode'' that works in case the GPSR fails to catch the GPS signals for some reasons. In this mode, called the ``Suzaku mode'', the SMU clock is no longer synchronized to the TAI and runs in the free-run mode\cite{Kouzu2011}, where the clock frequency changes depending on the temperature. In this case, the time drift due to this frequency change has to be corrected in converting the TIs to the TIMEs.

Considering the scientific needs, the requirement of the absolute time accuracy is set to be 1 ms (at the 1 $\sigma$ level) for the entire XRISM timing system, regardless of whether or not the SMU clock is synchronized to the TAI (see Terada et al. 2025\cite{Terada2024JATIS}). The error budgets are assigned to the individual components forming the timing system. An error budget of 350 $\mu$s is assigned to the spacecraft bus system (including the SMU) and the ground data processing. We refer the readers to Terada et al. (2025)\cite{Terada2024JATIS} for more details of all individual error budgets. We have conducted ground and on-orbit verifications to confirm that this requirement is satisfied. In this paper, we focus on the verification of the Suzaku mode and report the results. The verification of the GPS synchronized mode is described in Terada et al. (2025)\cite{Terada2024JATIS}. In the following sections, we first explain how the time assignment is performed in the Suzaku mode (Section~\ref{sec:method_timassign}), and then describe the results of the ground tests, mainly those from the thermal vacuum test conducted in 2022 (Section~\ref{sec:ground}) and the results from the on-orbit commissioning conducted within $\sim$ 1 month after the launch (Section~\ref{sec:orb}). Part of these results have been covered in Shidatsu et al. (2024)\cite{Shidatsu2024SPIE}. Section~\ref{sec:summary} gives the summary. Note that the SMU has identical two components, the SMU-A and the SMU-B, for redundancy. In the ground tests, we investigated both the two components, but only the SMU-A has been operated after the launch and was used in the on-orbit commissioning. In this paper, we omit the SMU-B results in the ground tests, which were found to be almost the same as those of the SMU-A.

\section{Time Assignment in the Suzaku Mode}
\label{sec:method_timassign}

In the Suzaku mode, the TI-to-TIME conversion and the assignment of the TIME values to all the satellite data are performed in the ground system following the same procedure as Hitomi\cite{Terada2018}. These time conversion and assignment processes are applied in the pre-pipeline processing\cite{Terada2021} using the dedicated HEAsoft tool {\tt xamktim} and {\tt xatime}. {\tt xamktim} produces the TI versus TIME data list (called the TIM FITS file; see below) referring to the bus-system housekeeping data containing the SMU-related information, and {\tt xatime} assigns the TIME values of the individual data using the list made in {\tt xamktim} as a look-up table between TI and TIME.

As mentioned above, the SMU generates TIs using its free-run clock in the Suzaku mode so the actual time of the TI interval varies with the frequency of the quartz oscillator in the clock, which changes depending on its temperature. To monitor this variation and convert TI to TIME accurately, a look-up table listing accurate TI versus TIME combinations, which is called the time calibration table, is created from the timing information contained in the ``time packet'' telemetry. This information is stored in the ``TIME\_PACKETS'' extension of TIM FITS file.
The time packet telemetry is generated in the telemetry command interface module (TCIM), which latches the TI counter in $\mu$s resolution every time it sends out the transfer frame to the ground station.
The ground station that downlinked the telemetry records the arrival time of the transfer frame in TAI, and then the TAI value is corrected for the delay from the time when TCIM latches the TI to the time when the ground station 
acquires the arrival time (such as the propagation delay between the satellite and the ground station),
all of which have known fixed values or can be calculated in the ground system very precisely. This realizes to obtain an accurate TAI value corresponding to the TI, with an error within $\sim$ 10 $\mu$s and a jitter of $\sim 1$ $\mu$s\cite{Terada2024JATIS}.

However, the time packet data (i.e., TIME\_PACKETS extension of TIM FITS file) can be obtained only during the communication between the satellite and a ground station, and especially during the period when the satellite is not visible from any ground stations, the data interval becomes too large to define the TIME values with the required accuracy. In the actual on-orbit operation, the time calibration table is created from the time packet data obtained 
at the ground station at Uchinoura Space Center (USC) in Kagoshima, Japan, and JAXA Ground Network (GN) stations in other countries (the Maspalomas station in Spain, the Mingenew station in Australia, and the Santiago station in Chile). 
Note that the time packet data taken at USC was only used and those at the GN stations were not used for the source of the time calibration table in the Hitomi case.
The communication is conducted $\sim 10$ times per day with a duration of each communication of typically $\sim 10$ minutes, and the data gap in the time calibration table becomes as large as 1--3 hours. When only the ground station at USC, from which the satellite is visible $\sim$ 5 sequential orbits (with an interval of $\sim 90$ minutes) per day, is available, the maximum data gap can be $\sim$ 2/3 of the day. During the communications, the time packet data are generated every 30 s, and only the first and last data in each communication are usually used for the time calibration table and stored in the TIME\_PACKETS extension of TIM FITS file. 

To obtain a TI versus TIME look-up table with sufficient sampling intervals for scientific purpose, we use the temperature dependence of the quartz frequency measured in advance and the actual SMU-temperature data (normally collected at 1-second intervals in orbit) and estimate the short-term variation of the TI-TIME relation between the accurate data points provided by the time calibration table. The temperature versus frequency trend was measured in ground tests before the launch (Section~\ref{subsec:ground_fvt}) and stored in the XRISM Calibration Database (CALDB). To examine its changes over time, the trend is also monitored in orbit about once every six months including the commissioning phase (Section~\ref{sec:orb}), and the CALDB files are updated if necessary. The quartz frequency is measured in the SMU when it is in the GPS synchronized mode, by comparing the 1PPS signals from the GPSR and the raw clock counter signals from the free-run clock before latched with the GPS signals. This measurement is carried out by sending a dedicated command from the ground station to the satellite. The temperature around the clock is always monitored even when the clock frequency measurement is not conducted. These data are stored in the bus-system housekeeping data file, from which the temperature versus frequency trend is derived in the ground system using the HEAsoft tool {\tt xatrendtemp}.

      \begin{figure} [th]
   \begin{center}
   \includegraphics[width=17cm]{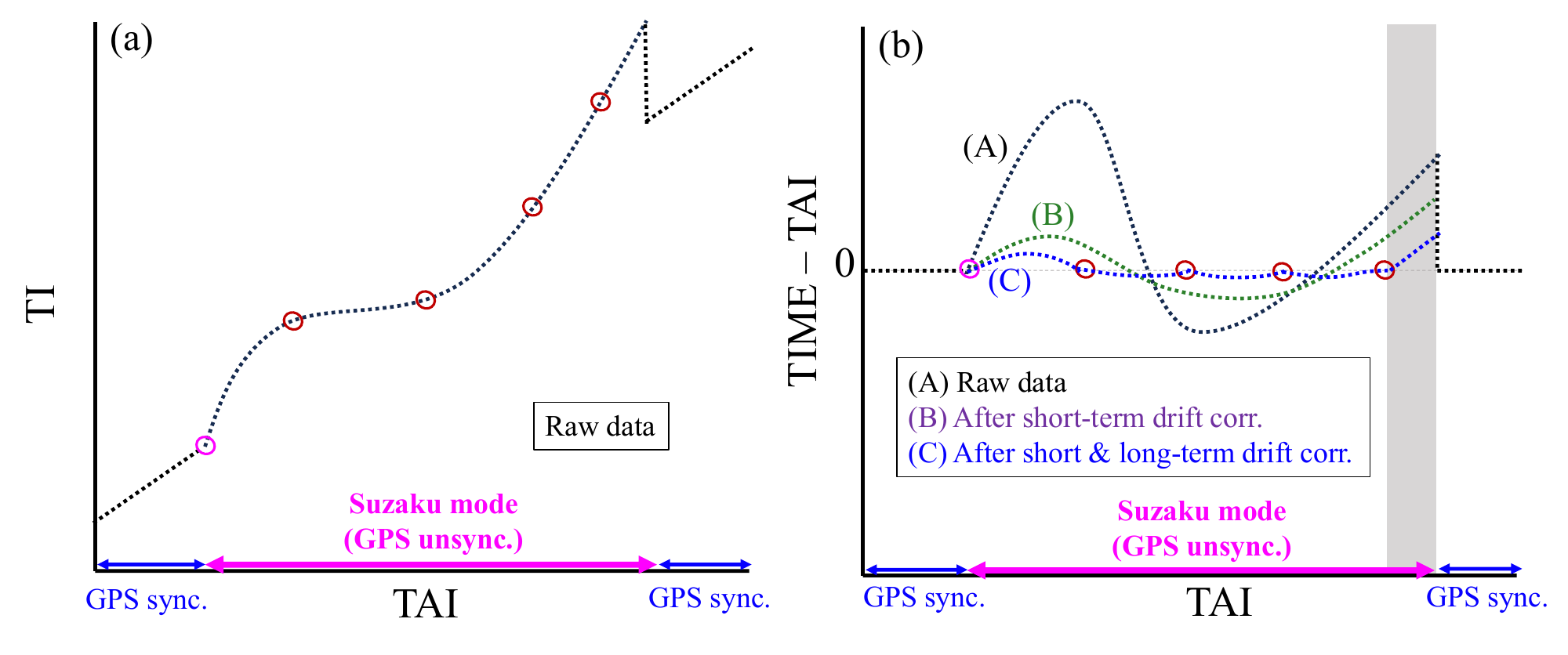}
	\end{center}
   \caption[Schematic view of the calculation of TIME in {\tt xamktim}.] 
   {\label{fig:xamktim} Schematic view of the calculation of TIME in {\tt xamktim} when the SMU switches from the GPS synchronized mode to the Suzaku mode and then back to the GPS synchronized mode. (a) The relation between TAI and TI. (b) TAI versus TIME $-$ TAI at different stages in the calculation. The black dotted line (A) adopts the TIME converted from the TI assuming the SMU is always synchronized to the TAI. The green (B) and blue (C) dotted lines show the TIME $-$ TAI value after the short-term time drift correction is performed and the both short and long-term corrections are performed, respectively. The pink and red circles show the last point of the GPS synchronized mode and the time calibration table data, respectively, which are used as anchor points for the long-term time drift correction. The axes of panels (a) and (b) are in arbitrary units.}
   \end{figure} 

The finer look-up table between TI and TIME is produced with {\tt xamktim} in the following procedure using the bus-system housekeeping data, as schematically shown in Figure~\ref{fig:xamktim}, and stored in the TIM\_LOOKUP extension of the TIM FITS file (Figure 3 of Terada et al. 2017\cite{Terada2018}). First, the clock frequencies at the individual TIs in the data are estimated from the corresponding SMU temperatures, referring to the temperature dependence of the quartz frequency in the CALDB. Using these frequencies, the short-term time drift is corrected for the individual periods between two adjacent data points. In this way, we obtain the TI versus TIME relation during the Suzaku mode period (B in Fig.~\ref{fig:xamktim}). Then, to reduce the remaining errors, the long-term correction is performed 
using the time calibration table data in the TIME\_PACKETS extension in the TIM FITS file as anchor points (C in Fig.~\ref{fig:xamktim}). 
In this correction, the Suzaku mode period is divided into segments using the time calibration table data and the slope and offset of the TI versus TIME relation for each segment are corrected so that it is smoothly connected to the leading and trailing anchor points. The last data point in the GPS synchronized mode before switching to the Suzaku mode is also used as the first anchor point. 
Through the above procedure, we finally obtain a look-up table between TI and TIME at $\sim$ 1-s intervals.

When the GPSR receives the GPS signals again, the SMU switches from the Suzaku mode to the GPS synchronized mode, through the transition mode. The transition mode lasts for about a few ten seconds, during which the TI is synchronized to the TAI. Even if the input data include both the GPS synchronized period and the Suzaku mode period, {\tt xamktim} determines the status of GPS synchronization using the bus system housekeeping data, calculate TIMEs from the TIs in an appropriate manner, and produce a TI versus TIME look-up table, which is stored in the TIM\_LOOKUP extension in the TIM FITS file. 
Note, however, that the long-term drift correction in the Suzaku mode cannot be applied for the period from the last time calibration data point in that mode to the start of the GPS synchronized mode (the shaded region in Fig.~\ref{fig:xamktim}), because the TI behaves in an unknown way during the transition to the GPS synchronized mode and the trailing anchor point is unavailable. This period is not the scope of the timing verification.
Finally, the TI-to-TIME conversion and assignment of the TIME values are carried out for all the satellite data with {\tt xatime} through linear interpolation of the two neighboring data points in the look-up table created with {\tt xamktim}.

\section{Ground Verification}
\label{sec:ground}

We have conducted the ground tests for timing verification in 2021 and in 2022. The tests in 2021 were the sub-system level tests and conducted at a test facility in NEC Corporation from January 26 to 28 and at JAXA Tsukuba Science Center (TKSC) on September 13. The test in 2022 was the thermal vacuum test conducted in the fully integrated spacecraft configuration using the thermal vacuum test chamber at TKSC in August 4--31. 
The main purpose of these tests was to confirm that the requirement for the absolute time accuracy (within an error of 350 $\mu$s) for the bus system and the ground data processing is satisfied both in the GPS synchronized mode and in the Suzaku mode. We also measured the SMU clock frequencies at different temperatures and studied their relation in both the sub-system level test and the thermal vacuum test. 

      \begin{figure} [ht]
   \begin{center}
   \includegraphics[width=12cm]{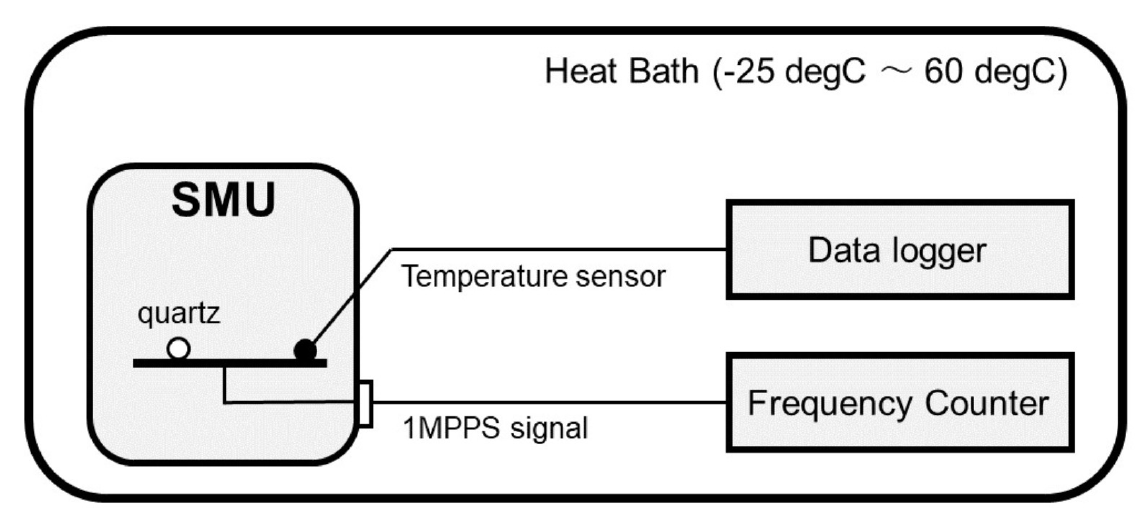}
	\end{center}
   \caption[Schematic picture of the configuration in the SMU temperature and the clock frequency measurement in the sub-system level test.] 
   {\label{fig:config_fvt} Schematic picture of the configuration in the SMU temperature and the clock frequency measurement in the sub-system level test.}
   \end{figure}

   \begin{figure} [hbt]
   \begin{center}
   \begin{tabular}{cc} 
   \includegraphics[height=8cm]{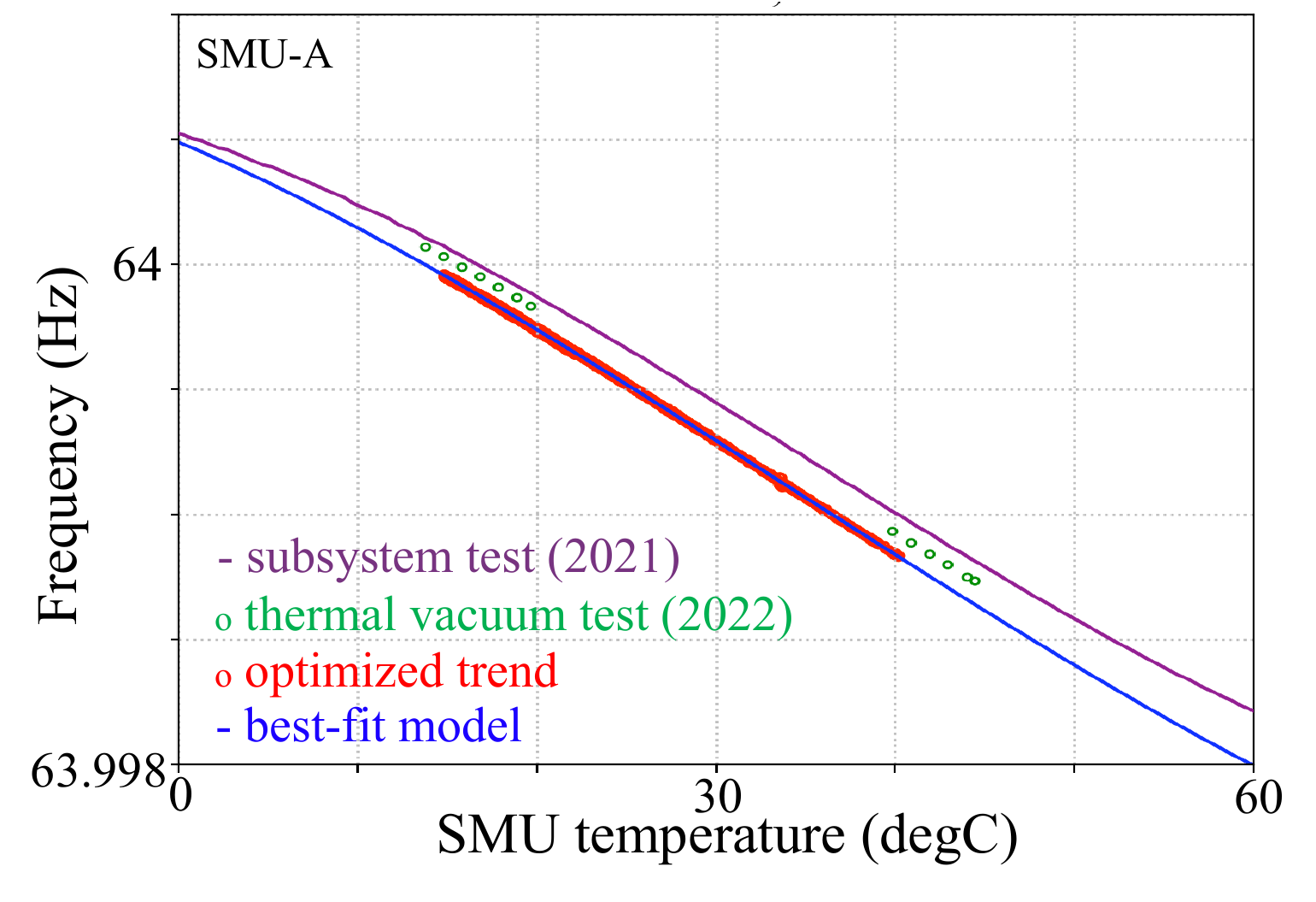} 
	\end{tabular}
	\end{center}
   \caption[Clock temperature versus frequency trends.] 
   {\label{fig:trendtemp} The temperature versus frequency trends of the SMU-A obtained in the sub-system level test (purple line) and the thermal vacuum test (green circles). See Sec.~\ref{sec:method_timassign} for the details of the method of measurement. The optimized trend (red circles; see Sec.~\ref{subsubsec:time_assign}) and its best-fit model of a third polynomial function (blue line) are also plotted.}
   \end{figure} 

\subsection{Temperature Dependence of the Clock Frequency}
\label{subsec:ground_fvt}

In Figure~\ref{fig:config_fvt} we present the schematic picture of the experimental setup in the temperature and clock frequency measurement in the sub-system level test. The measurements were conducted in the heat bath at the atmospheric pressure. The frequency of the quartz was measured by a frequency counter using the 1 Mega pulse per second (MPPS) signals generated from the SMU. The temperature was measured using a temperature sensor on the electric board on which the quartz was mounted. The temperature of the heat bath was varied between $-$25 degC and 60 degC several times. In the thermal vacuum test, the measurement was carried out at $\sim 10$ different temperatures in the fully integrated spacecraft configuration. 

Figure~\ref{fig:trendtemp} shows the trends in the frequency of the SMU clock with respect to the temperature, measured in the sub-system level test and thermal vacuum test. Note that the measurement was conducted from $-$25 degC to 60 degC in the sub-system level test but we focused on the results above 0 degC in Fig.~\ref{fig:trendtemp}, considering the actual temperature range (around 20--30 degC, see Section~\ref{sec:orb}) expected in orbit. The two tests gave almost consistent, but slightly different trends. The minor systematic differences may arise because the temperature measurement in each test was done at different points on the SMU.

\subsection{Verification of Time Assignment Using Thermal Vacuum Test Data}
\label{subsec:ground_tim}

\subsubsection{Test Configuration}

      \begin{figure} [ht]
   \begin{center}
   \includegraphics[width=16cm]{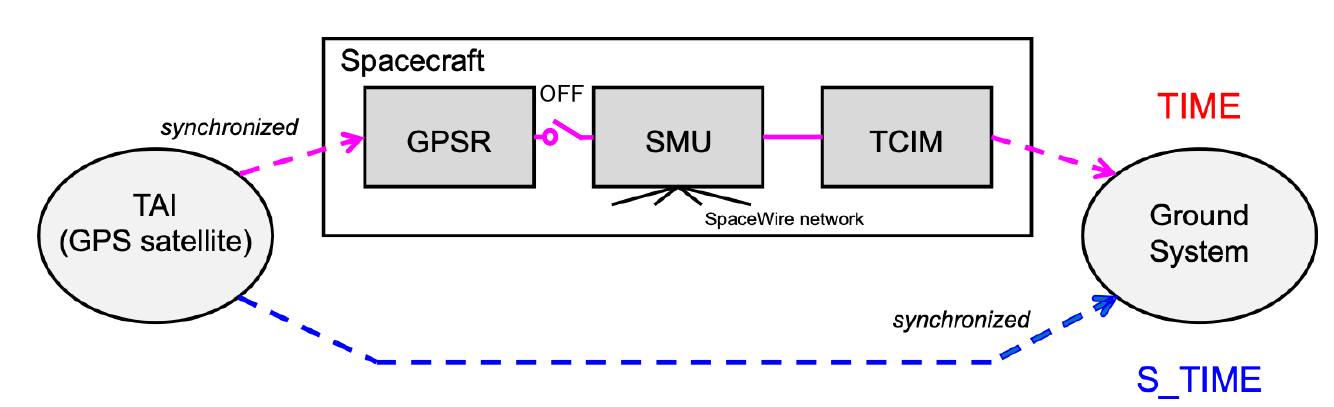}
	\end{center}
   \caption[Schematic picture of the test configuration in the Suzaku mode.] 
   {\label{fig:config} Schematic picture of the test configuration in the Suzaku mode, in which the SMU clock was not synchronized to the TAI. The ground system is a full emulator of the ground telemetry command communication equipment, which uses the reference clock synchronized to the GPS time.
    Time packets from the telemetry generated in the TCIM (see Sec.~\ref{sec:method_timassign}) were analyzed for verification.}
   \end{figure} 

In Figure~\ref{fig:config} we show the schematic picture of the test configuration of timing verification for the Suzaku mode. The ground system was always synchronized to the TAI, while the SMU clock was operated in the free-run mode, which was realized by artificially switching off the connection between the GPSR and the SMU by a dedicated command. The TI values generated by the SMU clock were sent as telemetry data to the ground system and used to calculate the TIME values. The actual TAI values at which the data were sent to the ground system were stored as S\_TIME, which can be used to study the accuracy of the TIME values. 
For the test configuration in the GPS synchronized mode, 
see Figure~1 of Terada  et al. (2024)\cite{Terada2024JATIS} . 

\subsubsection{Time Assignment Test}
\label{subsubsec:time_assign}
We performed time assignment of data obtained in the thermal vacuum test and investigated the error in the absolute time by comparing the TIME values converted from the TIs to the actual TAI values (S\_TIME), following the procedure described in Section~\ref{sec:method_timassign}. In this test, we adopted the time packet data 
for the combinations of the TI versus S\_TIME values used in the time assignment, 
instead of the housekeeping or event data of the bus-system and the instruments, produced from normal spacepackets. The time packet data were generated every 30 s, as in the actual on-orbit operation. As mentioned in Section~\ref{sec:method_timassign}, 
the time packet data are used as the source to create the time calibration table and provide accurate timing information with small timing jitters (of order 1 $\mu$s). By contrast, the data from normal spacepackets contain large jitters and delays in the S\_TIME values (of order $\sim 10$ ms) that cannot be removed, and hence are not suitable for this test. The large jitters and delays in normal spacepackets  are because, unlike the time packets, the normal spacepackets do not involve the special process of latching the TI value when generating them (see Section~\ref{sec:method_timassign} for details). We compared the time packet data and the bus-system housekeeping data, to extract the other information required in the time assignment (such as flags to determine the GPS synchronization status and the SMU temperature) from the latter data.

   \begin{figure} [ht!]
   \begin{center}
   \begin{tabular}{cc} 
   \includegraphics[width=12cm]{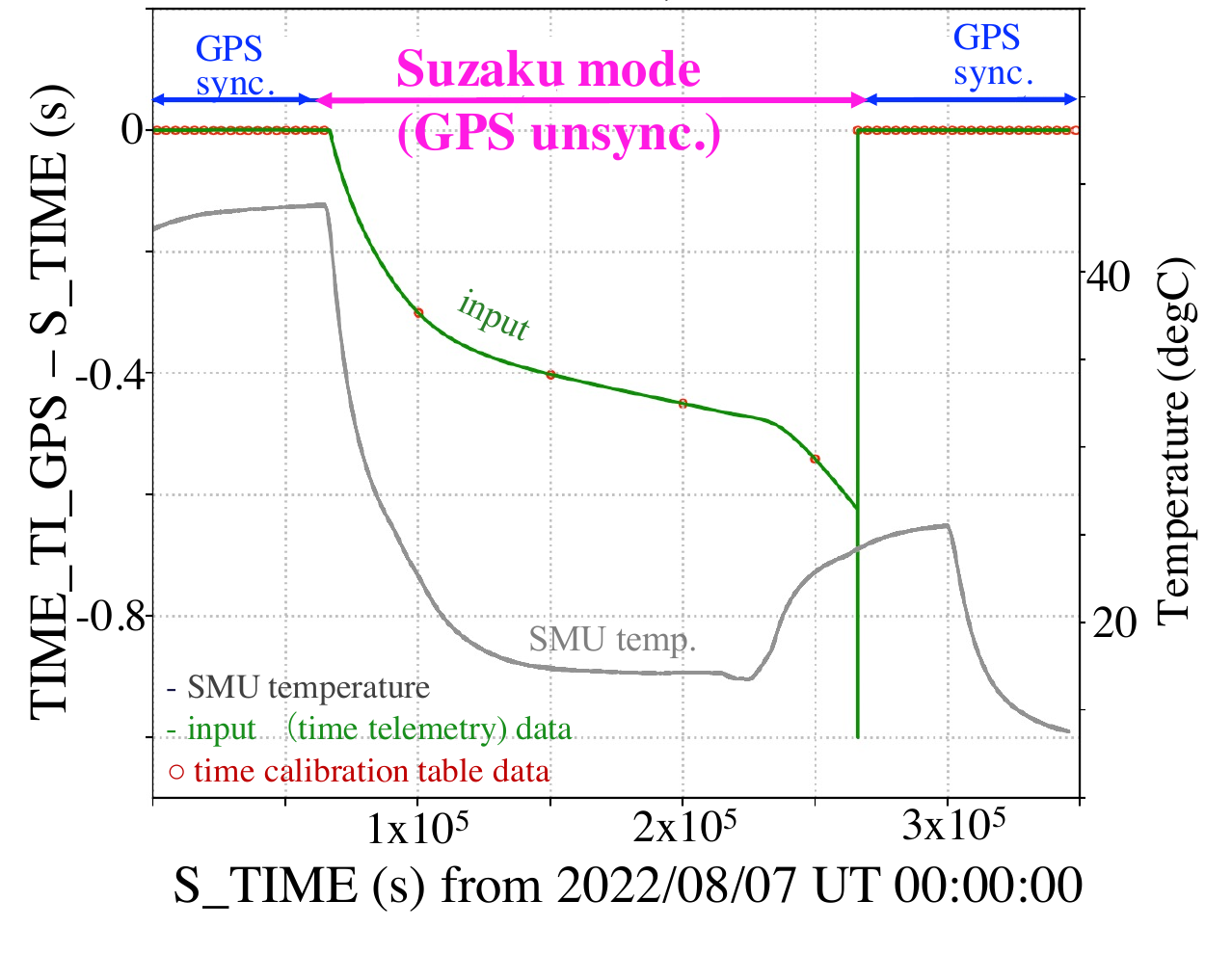} 
	\end{tabular}
	\end{center}
   \caption[Time variation of the TI and the SMU temperature in the thermal vacuum test.] 
   { \label{fig:input_tvt} 
Variations of the time packet data used for the timing verification in the thermal vacuum test (green) and of the SMU temperature (gray) in the SMU-A. TIME\_TI\_GPS in the vertical axis is the ``TIME'' value converted from the TI assuming the GPS synchronized mode. The large ($-1$ s) jump in TIME\_TI\_GPS $-$ S\_TIME at the end of the Suzaku mode period is a normal behavior of the SMU at the transition to the GPS synchronized mode. The red circles present the time calibration table data used for the long-term drift correction in conversion from TI to TIME.}
   \end{figure} 
   
  \begin{figure}[hbt!]
   \begin{center}
   \begin{tabular}{cc} 
   \includegraphics[height=10cm]{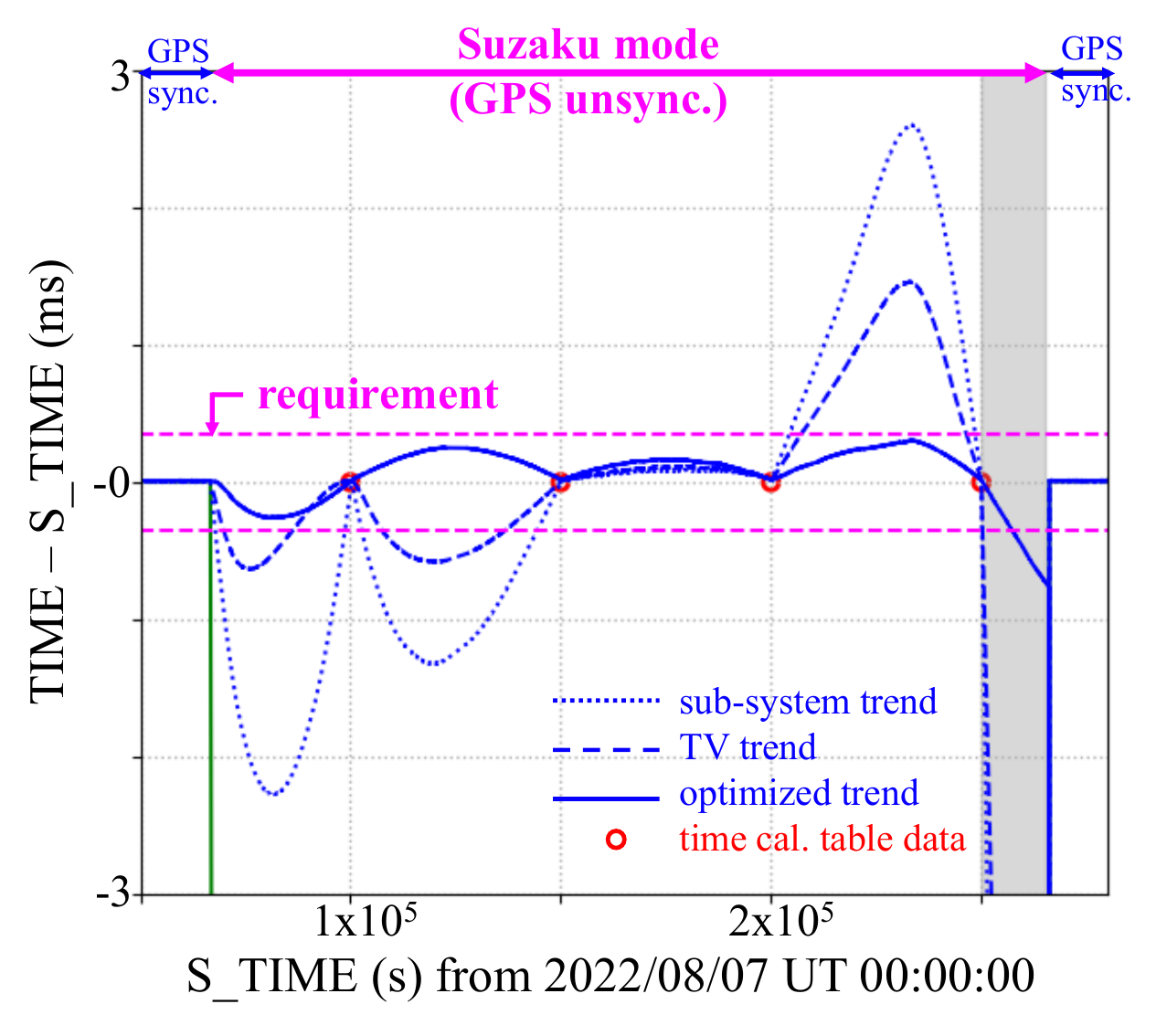} 
	\end{tabular}
	\end{center}
   \caption[Results of time assignment of the data obtained during the Suzaku mode period in the thermal vacuum test.] 
   { \label{fig:result_tvt} 
The results of the time assignment of the time packet data. The vertical axis shows the deviation of the resultant TIME value from the actual time (S\_TIME). The dotted, dashed, and solid lines show the results using the temperature-versus-frequency trend obtained from the sub-system level test, that from the thermal vacuum (TV) test, and the optimized trend, respectively. The horizontal magenta dashed lines indicate the requirement of the timing error (i.e., $\pm 350$ $\mu$s). The green, almost vertical line at S\_TIME $= 7 \times 10^4$ s shows the input data (same as Fig.~\ref{fig:input_tvt}). Note that the shaded region, from the last time calibration data point in the period of the Suzaku mode to the start of the GPS synchronized mode, is not the scope of the timing verification (see Sec.~\ref{sec:method_timassign}). 
}
   \end{figure}

Figure~\ref{fig:input_tvt} shows the input data made from the time packet data (green) and the trend in the SMU temperature (grey). In this plot, the TI values were converted to the ``TIME'' values, assuming that the clock had always been in the GPS synchronized mode (TIME\_TI\_GPS in the figure). TIME\_TI\_GPS deviated from the S\_TIME after switching from the GPS synchronized mode to the Suzaku mode, due to the change in the clock frequency. 
We made the time calibration table data from the same time packet data, for the long-term drift correction in the TI-to-TIME conversion. Although the time packet data were generated with an interval of 30 s, we selected 
one data point every $5 \times 10^4$ s, which corresponds to the maximum interval ($\sim$ 2/3 day) between the communication in successive orbits when JAXA GN stations are unavailable,
and discarded the remaining data points.

Figure~\ref{fig:result_tvt} presents the results of the time assignment of the time packet data. The error in TIME was found to be larger than the requirement (350 $\mu$s) when we adopted the temperature-versus-frequency trend from the sub-system test. The situation did not change when the trend from the thermal vacuum test was applied, although the error was somewhat reduced relative to 
the previous case.

The results suggest that the measured frequency versus temperature trends do not have sufficient accuracy. Then, we used the sets of the TI and S\_TIME values of the time packet data and derived the best temperature versus frequency trend that makes the TIME values converted from the TIs always consistent with the S\_TIME values. 
To produce this trend, we first calculated the differences in TI and in S\_TIME (which is the actual time interval in the TAI) for each of the two neighboring data points in the Suzaku mode. Then, we estimated the frequency at each time interval so that the difference of TI times the frequency equals the difference in S\_TIME (i.e., $f = \Delta {\rm S\_TIME} / \Delta {\rm TI}$, where $f$ is the frequency, and $\Delta {\rm S\_TIME}$ and $\Delta {\rm TI}$ are the differences of S\_TIME and TI, respectively). In addition, we derived the average of the temperatures measured at those two points and adopted them as the clock temperature at that time interval. Finally, the frequency and the temperature combinations obtained in this procedure were fitted with a third polynomial function and the best-fit model was adopted as the ``optimized'' trend. 

The resultant dataset of the frequency and the temperature and the optimized trend of the SMU-A is plotted in Figure~\ref{fig:trendtemp}, which has a small offset from the measurements in the sub-system level test and the thermal vacuum test. This deviation may be produced by the uncertainty in the temperature that is not measured at the quartz itself but a different point on the SMU. The results of the time assignment using the optimized trend are plotted in Figure~\ref{fig:result_tvt}. The errors in the TIME values were found to be within the requirement. Note that errors were not completely zero even when we use the optimized trend. This is likely because the time packet data were generated in the intervals of about 30 seconds, during which the temperature changed slightly. This temperature change could affect the trend as an uncertainty and produce errors in the time assignment.

\subsection{Simulation of Timing Accuracy in On-orbit Conditions}
\label{subsec:ground_sim}

The verification in Section~\ref{subsec:ground_tim} was performed in a special temperature condition adopted in the thermal vacuum test. Here, we investigated how the timing accuracy changes in the typical on-orbit temperature condition. 
For this purpose, we used Hitomi on-orbit data as a reference. XRISM adopts the same bus system as the Hitomi spacecraft and operates in almost the same low-Earth orbit\cite{Tashiro2025}, so similar temperature conditions are expected to be realized in the SMU. The SMU clock also uses the same hardware as that on Hitomi, and the crystal oscillator of Hitomi has a similar temperature versus frequency trend as XRISM\cite{Terada2018}. For these reasons, Hitomi on-orbit data serve as a suitable reference for evaluating the XRISM timing accuracy in on-orbit conditions.

  \begin{figure} [ht]
   \begin{center}
   \includegraphics[height=5.cm]{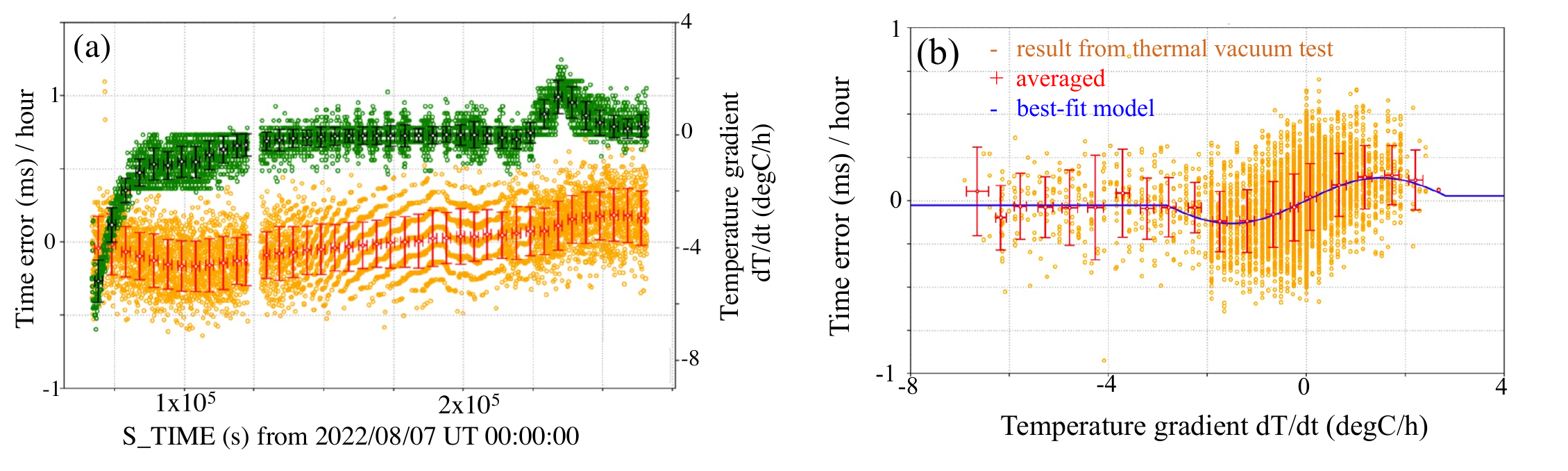}
	\end{center}
   \caption[Dependence of the timing errors on the temperature gradient.] 
   {\label{fig:tvt_gradT_error} (a) Time evolution of the gradient of the SMU temperature (green) and the timing accuracy estimated from the time packet data (orange). The black and red data show the same data, but binned with each 5000 s interval. The errors represent the standard deviation. Note that the long-term correction was ignored in the timing accuracy. (b) Dependence of the timing accuracy on the temperature gradient. Orange and red points show the original data and those binned with a $dT/dt =$ 0.5 degC/h interval. The error bars of the orange data represent the standard deviation. The blue line presents a model to characterize the orange data.}
   \end{figure}

First, we studied the dependence of the timing accuracy on the temperature gradient, using the results of time assignment in Section~\ref{subsec:ground_tim}, in the case of the optimized temperature versus frequency trend. Figure~\ref{fig:tvt_gradT_error}(a) presents the time evolution of the temperature gradient and the timing accuracy, and Figure~\ref{fig:tvt_gradT_error}(b) shows their correlation. In this plot, we estimated the timing accuracy considering only the short-term time drift correction, without the long-term correction using the time calibration data.

\begin{figure} [ht]
   \begin{center}
   \includegraphics[height=5.5cm]{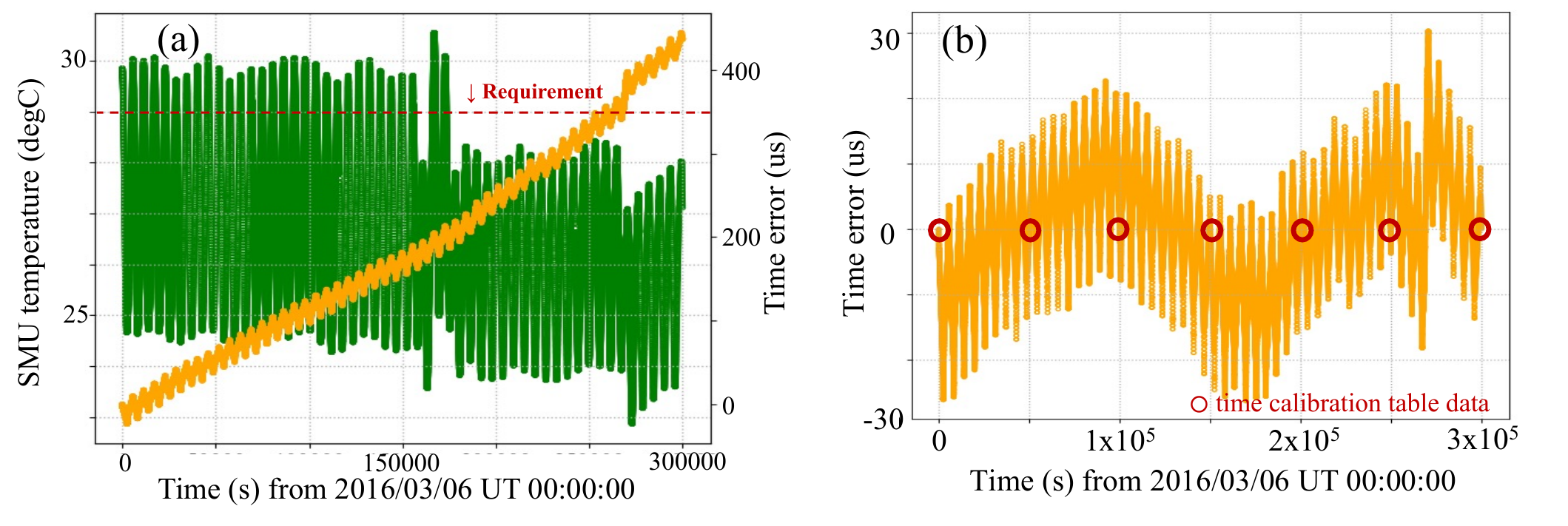}
	\end{center}
   \caption[Simulation of the XRISM timing error using the Hitomi on-orbit data of the SMU temperature.] 
   { \label{fig:sim} 
(a) The SMU temperature of Hitomi measured on-orbit (green) and the variation of the XRISM timing error (orange) estimated using the Hitomi temperature data and the correlation between the temperature gradient and the timing error obtained in Figure~\ref{fig:tvt_gradT_error}. 
(b) The timing error obtained in (a), in which the long-term time drift correction was performed using the time calibration data (red circles) with an interval of $5 \times 10^4$ s.}
   \end{figure}

Next, we estimated 
temporal evolution of timing inaccuracy using this correlation and the Hitomi on-orbit data of the SMU temperature. Figure~\ref{fig:sim}(a) plots the results in a typical case of the temperature variation. The error was found to remain within the requirement for $\sim 2.5 \times 10^5$ s. If we use the time calibration table data with an interval of $5 \times 10^4$ s to long-term drift correction, the requirement is satisfied for at least $\sim 3 \times 10^5$ s in typical temperature conditions of Hitomi (Figure~\ref{fig:sim}). This duration is comparable with the Suzaku case\cite{Terada2008}. This also satisfies the requirement of the duration (17 hours $= 6.12 \times 10^4$ s) for which the error in the absolute time should be kept within the 350 $\mu$s error budget.

\section{On-orbit Commissioning}
\label{sec:orb}

After the launch, we conducted on-orbit commissioning of the Suzaku mode. The commissioning activities were done for the first eight days after the launch and about one month later, both of which were carried out as part of the bus system checkout. The main purpose of these activities was to 
test the functionality to switch between the Suzaku mode and the GPS synchronize mode and confirm that
the SMU clock in the Suzaku mode produces proper TI values following the temperature versus frequency trend. Note that the GPSR has been receiving the GPS signals stably and the SMU has always been in the GPS synchronized mode as of 2025 April, except during these commissioning activities (i.e., the Suzaku mode has been active for $\sim 1$\% of the total operational time on orbit, all of which was intentionally enabled).

\begin{figure} [ht]
   \begin{center}
   \includegraphics[height=7cm]{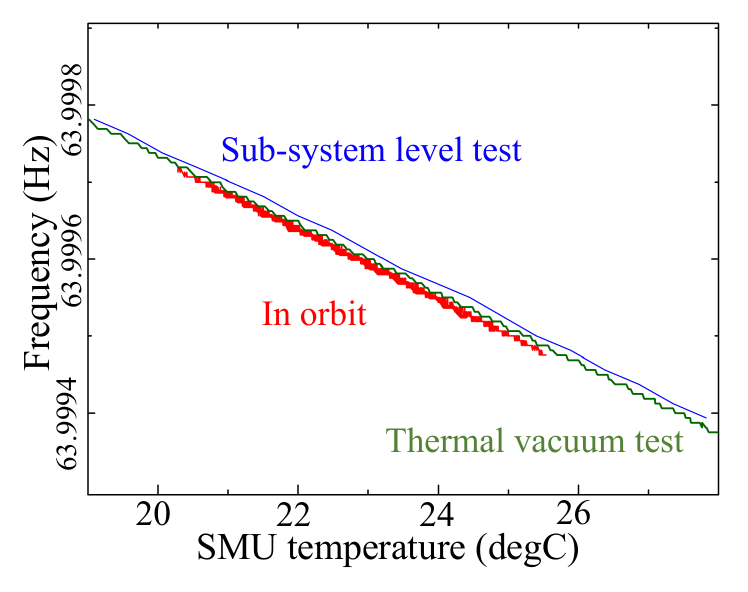} 	
   \end{center}
   \caption[Temperature versus frequency trend measured in orbit] 
   { \label{fig:onorb_trendtemp} 
The temperature versus frequency trend of the SMU-A measured in orbit (red). See Sec.~\ref{sec:method_timassign} for the details of the method of measurement. The trend obtained in the sub-system level test (blue) and the thermal vacuum test (green) are also plotted. 
}
   \end{figure}

Figure~\ref{fig:onorb_trendtemp} shows the temperature versus frequency trend of the SMU-A obtained in orbit, compared with those from the ground tests. The on-orbit trend is consistent with the one that was measured in the thermal vacuum test, suggesting that the properties of the SMU clock have not changed significantly after the launch. The temperature and clock frequency measurement will be conducted about once per half year 
to monitor any temporal variation of the trend in orbit.

 \begin{figure} [ht]
   \begin{center}
   \includegraphics[height=5.8cm]{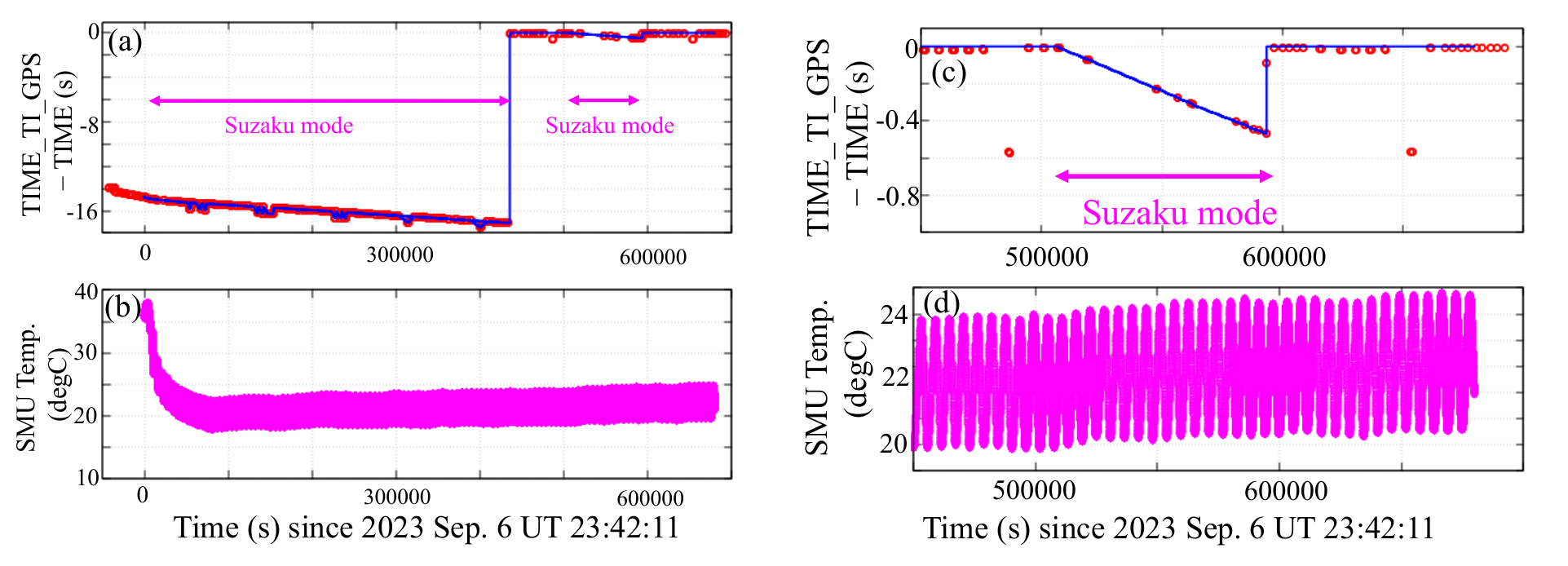}
	\end{center}
   \caption[Results of the on-orbit timing verification of the Suzaku mode] 
   { \label{fig:onorb} 
(a) The time variation of the TI and TIME values taken in the initial eight days after the launch. The vertical axis presents the difference between the TIME\_TI\_GPS values (converted from TI assuming the GPS synchronized mode) and the actual TIME values assigned to the data. The red circles present the time calibration data. (b) Variation of the SMU temperature in the same period as (a). (c),(d) An enlarged view of (a) and (b) around the 6th--7th day from the launch, respectively. 
}
   \end{figure}

Figure~\ref{fig:onorb} plots the time variation of the TIME\_TI\_GPS $-$ TIME values and the SMU temperature during the initial eight days after the launch, obtained from the time packet data and the bus-system housekeeping data, respectively. The SMU clock was in the Suzaku mode at the launch and the GPS synchronization functions were established 5 days ($4.4 \times 10^{5}$ s) after the launch. The clock was again switched to the Suzaku mode and then back to the GPS synchronized mode in the 6th to 7th days. 
 Adopting a typical temperature of $\sim 21$ degC during these periods and using the temperature versus frequency trend (Fig~\ref{fig:trendtemp} or Fig.~\ref{fig:onorb_trendtemp}), the integrated time drifts are estimated to be $\sim 1.6$ s and $\sim 0.47$ s during the Suzaku mode periods before the establishment of the GPS synchronized mode and in the 6th--7th days, respectively. These values are consistent with the actual change in TIME\_TI\_GPS $-$ TIME observed in these two periods, indicating that the timing system in the Suzaku mode works properly in orbit.

\section{Summary}
\label{sec:summary}
It was not straightforward to satisfy the requirement for timing accuracy in the Suzaku mode. In the thermal vacuum test, we found that the measured temperature versus frequency trend has insufficient accuracy to use in the time assignment. We optimized the temperature versus frequency trend using the time packet data and successfully achieved sufficient accuracy with errors of $\leq 350$ $\mu$s. Further investigation of the timing errors assuming a more realistic temperature condition suggests that the requirement is expected to be satisfied for at least $3 \times 10^5$ s. The results from the on-orbit commissioning indicate that the timing system in the Suzaku mode keeps working properly after the launch. 

We were able to demonstrate that the timing accuracy in this mode meets the requirement. 
 However, this verification is limited to the pre-launch period and a relatively short duration after launch. Although the system is designed to maintain the required accuracy throughout the planned three-year mission, its performance may gradually degrade over time due to e.g., the radiation environment in orbit and long-term aging effects. As mentioned above, we will continue to monitor potential performance degradation through temperature and clock frequency measurements performed every $\sim$ 6 months.

%\acknowledgments % equivalent to \section*{ACKNOWLEDGMENTS}       

%\appendix    % this command starts appendixes
%\section{Appendix title}
%\label{sect:app1}

\subsection* {Disclosures}
The authors declare that there are no financial interests, commercial affiliations, or other potential conflicts of interest that could have influenced the objectivity of this research or the writing of this paper.

\subsection* {Code, Data, and Materials Availability}
The data that support the findings of this article are proprietary and are not publicly available. The data plotted in the above figures are available from the corresponding author upon request, and a limited subset of the underlying data can be requested from the author at shidatsu.megumi.wr@ehime-u.ac.jp.

\subsection* {Acknowledgments}
 The authors thank the editor, Masanobu Ozaki, and the anonymous referees for providing valuable comments. The authors are grateful to the whole XRISM team for continuous efforts on developing, testing, and operating the spacecraft, its instruments, and the ground system. This work was supported by the JSPS Core-to-Core Program 
 (grant number: JPJSCCA20220002) and Grants-in-Aid for Scientific Research (KAKENHI) JP19K14762, JP23K03459, JP24H01812 (MS), and JP20K04009 (YT) from the Ministry of Education, Culture, Sports, Science and Technology (MEXT) of Japan. The work of ML, TY, KM, KH, and KP are supported by NASA under award number 80GSFC24M0006. EDM acknowledges support from NASA grants 80NSSC20K0737 and 80NSSC24K0678.

%%%%% References %%%%%

\bibliography{report}   % bibliography data in report.bib
\bibliographystyle{spiejour}   % makes bibtex use spiejour.bst

%%%%% Biographies of authors %%%%%

\vspace{2ex}\noindent\textbf{Megumi Shidatsu} is an associate professor at Ehime University. She received her BS, MS, and PhD degrees in science from Kyoto University in 2010, 2012, and 2015, respectively.  She is a member of the Science Operations team of the X-Ray Imaging and Spectroscopy Mission (XRISM) project and has been involved in the verification of the XRISM timing system and of the ground data processing ("pre-pipeline") software that includes the time assignment tools. She is also a member of 
the timing working group in the International Astronomical Consortium for High-Energy Calibration (IACHEC) and involved in  cross timing calibration among XRISM and different missions.

\vspace{2ex}\noindent\textbf{Yukikatsu Terada} is a professor at Saitama University and the Japan Aerospace Exploration Agency (JAXA). He received his BS and MS degrees in physics, and a PhD degree in science from the University of Tokyo in 1997, 1999, and 2002, respectively. He is the leader of the Science Operations Team of the X-Ray Imaging and Spectroscopy Mission (XRISM). He is chairing the timing working group in the International Astronomical Consortium for High-Energy Calibration (IACHEC).

\vspace{1ex}
\noindent Biographies and photographs of the other authors are not available.

\listoffigures
%\listoftables

\end{spacing}
\end{document}